\documentclass[aps,prl,twocolumn,groupedaddress,amsmath,amssymb]{revtex4-1} 
\usepackage{amssymb,amsfonts,amsmath}
\usepackage{graphicx}
\usepackage{amsbsy}
\usepackage{revsymb}

\usepackage{dcolumn}
\usepackage{bm}

\def\C{{\@QC C}}
\def\@QC#1{\mathpalette{\setbox0=\hbox\bgroup$\rm}%
  {\egroup C$\egroup\rm\rlap{\kern0.4\wd0\vrule
  width 0.05\wd0 height 0.97\ht0 depth -0.01\ht0}%
  #1\bgroup}}

\newcommand{\bc}{\begin{center}}
\newcommand{\ec}{\end{center}}
\newcommand{\be}{\begin{equation}}
\newcommand{\ee}{\end{equation}}
\newcommand{\bi}{\begin{itemize}}
\newcommand{\ei}{\end{itemize}}

\usepackage[latin1]{inputenc} 

\begin{document}

\title{Absence of Metallization in Solid Molecular Hydrogen}

\author{Sam Azadi}
\affiliation{Institute of Physical Chemistry and                                                  
Center for Computational Sciences, Johannes Gutenberg University
Mainz, Staudinger Weg 9, 55128 Mainz, Germany}
\author{Thomas D. K\"uhne}
\email{kuehne@uni-mainz.de}
\affiliation{Institute of Physical Chemistry and                                                  
Center for Computational Sciences, Johannes Gutenberg University
Mainz, Staudinger Weg 9, 55128 Mainz, Germany}

\date{\today}

\begin{abstract}
Being the simplest element with just one electron and proton the electronic structure of a single Hydrogen atom is known exactly. However, this does not hold for the complex interplay between them in a solid and in particular not at high pressure that is known to alter the crystal as well as the electronic structure and eventually causes solid hydrogen to become metallic. 
In spite of intense research efforts the experimental realization of metallic hydrogen, as well as the theoretical determination of the crystal structure has remained elusive.
Here we present a computational study showing that the distorted hexagonal P6$_3$/m structure is the most likely candidate for Phase III of solid hydrogen. 
We find that the pairing structure is very persistent and insulating over the whole pressure range, which suggests that metallization due to dissociation may precede eventual bandgap closure. Due to the fact that this not only resolve one of major disagreement between theory and experiment, but also excludes the conjectured existence of phonon-driven superconductivity in solid molecular hydrogen, our results involve a complete revision of the zero-temperature phase diagram of Phase III. 
\end{abstract}

\maketitle

Back in 1935 Wigner and Huntington \cite{JCP.3.764} predicted that at very high pressure solid molecular hydrogen would dissociate and form an atomic solid that is metallic. Due to its relevance to astrophysics \cite{1995Sci...269.1252A}, but in particular because of the possible existence of high-$T_{c}$ superconductivity \cite{PhysRevLett.21.1748} and a metallic liquid ground state \cite{2004Natur.431..669B}, the importance to grasp metallic hydrogen can hardly be overstated. 
Since it is by now still not possible to reach the necessary static compression ($>$ 400 GPa) to dissociate hydrogen, recently alternative routes to metallic hydrogen, but at lower pressure have been proposed. 
On the one hand, the negative slope of the melting line \cite{2003PNAS..100.3051S} immediately suggests the possibility of producing liquid metallic hydrogen at low finite temperature \cite{PhysRevLett.100.155701, 2009JETPL..89..174E, Eremets2011}. On the other hand, due to the persistence of the molecular phase, it has been predicted that even in the paired state metallization through bandgap closure may be possible \cite{PhysRevLett.34.812, PhysRevLett.62.1150}, which would be very consequential since it facilitates potential high-$T_{c}$ superconductivity in paired metallic hydrogen \cite{1989Natur.340..369B, PhysRevLett.78.118}. However, both avenues are complicated by the fact that contrary to Phase I ($<$ 110 GPa), which is the only quantum molecular solid and made of quantum rotors on a hcp lattice, the structures of Phase II and Phase III ($>$ 150 GPa) are still unknown \cite{PhysRevLett.61.857, PhysRevLett.63.2080}, in particular whether or not the latter is metallic \cite{1989Sci...244.1462M, PhysRevLett.76.1663, PhysRevLett.76.1667}. Since even the combined power of experimental vibrational and scattering data have not yet allowed for a unique crystal structure determination, only by means of theory a large variety of different structures have been predicted as potential candidates for Phase III, many of which were indeed metallic \cite{PhysRevLett.34.812, PhysRevLett.62.1150, PhysRevB.36.2092, PhysRevLett.66.64, PhysRevLett.67.1138, Alavi1998, PhysRevLett.83.4097, 2000Natur.403..632J, PhysRevLett.84.6070}.
Anyhow, because of the vast amount of different possible structures and the small energy differences among them, it is impossible to ensure that any structure represents the global minimum in enthalpy. 

However, recently great strides have been made to predict crystal structures from first-principles, to the extend that allowed Pickard and Needs to systematically investigate the zero-temperature phase diagram of Phase III of solid hydrogen \cite{2007NatPh...3..473P}. Specifically, for the pressure range of Phase III they predict manifold crystal structures to be energetically most favorable, namely up to 270 GPa the C2/c phase, followed by Cmca-12, before at 385 GPa solid molecular hydrogen eventually transforms into the Cmca phase. 

Neverthless, the determination of the metallization pressure in all the calculations based on the local or semilocal approximation to density functional theory (DFT) is plagued by the infamous bandgap problem that typically underestimates the true fundamental gap by as much as $\sim$50\% \cite{PhysRevLett.51.1884}. In fact, more accurate calculations based on many-body perturbation theory \cite{PhysRevLett.66.64}, as well as hybrid DFT \cite{PhysRevLett.84.6070} report on a substantial increase in the metallization pressure of hydrogen. Anyway, the considered structures have been shown to be energetically not competitive \cite{2007NatPh...3..473P}. Moreover, enthalpic effects because of potential bandgap corrections to the valence bands were omited, which otherwise would entail a stabilization of the insulating phases with respect to metallic ones and therefore further increase the metallization pressure. 



\begin{figure}
\centering
\includegraphics[width=1.0\columnwidth]{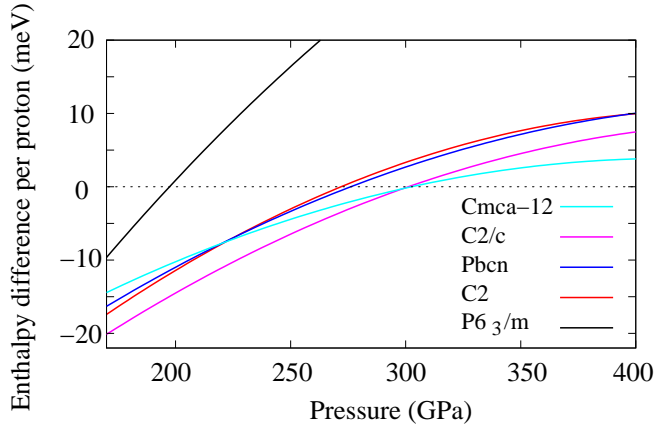}
\caption{
\label{enthalpy-pbe}
(Colors online) The enthalpy per proton relative to the metallic Cmca phase at the semilocal DFT level as a function of pressure.} 
\end{figure}

\begin{figure}
\centering
\includegraphics[width=1.0\columnwidth]{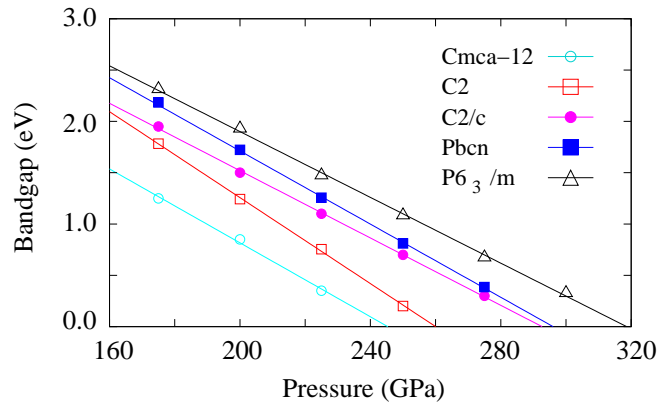}
\caption{
\label{PBE-gap}
(Colors online) The single-particle bandgap as functions of pressure, as obtained by semilocal DFT calculations.} 
\end{figure}

In this work, we revise the zero-temperature phase diagram of solid molecular hydrogen in the pressure range of Phase III using electronic structure methods that go beyond the semilocal DFT level of theory.
Specifically, we consider the C2/c, Cmca-12, C2, Pbcn, as well as the P6$_{3}$/m structures, which have been recently proposed and enthalpically found to be most favorable at the semilocal DFT level \cite{2007NatPh...3..473P}. Our DFT calculations were all carried out within the pseudopotential plane-wave approach using the Quantum Espresso suite of programs \cite{2009JPCM...21M5502G}. In order to accurately sample the Brillouin zone a dense {\bf k}-point mesh with at least $16^3$ special points were used so as to guarantee convergence of the total energy to 1 mRy per molecule. The very same criterion holds for the explicit minimization of the enthalpy for all considered structures and pressures by means of a concurrent geometry and cell optimization. 
For the calculations at the semilocal DFT level the Perdew-Burke-Ernzerhof (PBE) generalized gradient approximation \cite{PhysRevLett.77.3865} to the exact exchange-correlation functional was used together with the projector augemented wave method \cite{PhysRevB.50.17953} and a corresponding planewave cutoff of 60 Ry. The corresponding phase diagram is shown in Fig.~\ref{enthalpy-pbe} and is in excellent agreement with Ref. \cite{2007NatPh...3..473P}, except for an energy lowering by $\sim$8 meV of the metallic Cmca structure that renders the Cmca-12 phase unstable, which in any event is incompatible with experimental infrared (IR) spectra of Phase III of solid hydrogen \cite{1994Natur.369..384H}. 
Together with Fig.~\ref{PBE-gap}, which shows the single-particle gap with respect to pressure for all considered structures, it can be deduced that up to $\sim$290 GPa the C2/c phase is prevailing and still insulating. Even though, thereafter bandgap closure would in principle be observed, at the same time the C2/c structure simultaneously transforms into the metallic Cmca phase. The well known rule, "the lower the energy, the wider the gap" \cite{PhysRevLett.67.1138} does not apply here. For instance, the energetically rather competitive Cmca-12 structure would obey metallization because of bandgap closure starting from 245 GPa, while on the contrary the largest gap is observed for the P6$_3$/m phase, which at this level of theory is energetically not favorable at all. 
Until now, the bandgaps have been given only in terms of the single-particle Kohn-Sham (KS) gap. However, the exact KS gap differs from the true fundamental gap by the so called derivative discontinuity $\Delta_{\text{XC}}$ of the exchange-correlation potential \cite{PhysRevLett.51.1884}. Due to the fact, that $\Delta_{\text{XC}} = 0$ within local and semilocal DFT, the fundamental gap and thus the metallization pressure is often severely underestimated. Fortunately, Hartree-Fock exchange (HFX) allows us to approximate the exchange contribution of $\Delta_{\text{XC}}$, but owing to the nonlocality of the exchange potential generally overestimates the fundamental gap. 
As a consequence hybrid DFT, which includes only a small fraction of exact HFX, typically yields much improved bandgaps that are often in close agreement with experiment, even though it does not provide a general solution for the DFT bandgap problem.
In addition, due to the absence of an artificial self-repulsion between the occupied states, HFX not only exactly cancels the self-interaction contribution of the Hartree energy, but also energetically stabilizes insulating phases relative to metallic ones, which in general are sufficiently accurate described by standard DFT. For these reasons hybrid DFT is expected to substantially increase the eventual metallization pressure, by favoring insulating phases and concurrently predicting throughout larger bandgaps. For our hybrid DFT calculations we have employed the so called PBE0 exchange-correlation functional, where according to the adiabatic connection 25\% of PBE exchange has been substituted by exact HFX \cite{JCP.110.6158}. In all hybrid DFT calculations a rather hard norm-conserving pseudopotential was utilized in conjunction with a planewave energy cutoff of 90 Ry. The additional quasiparticle bandgap calculations at the many-body perturbation level of theory have been performed within the G$_{0}$W$_{0}$ approximation \cite{PhysRevLett.55.1418} using the Yambo code \cite{2009CoPhC.180.1392M}. 
\begin{figure}
\centering
\includegraphics[width=1.0\columnwidth]{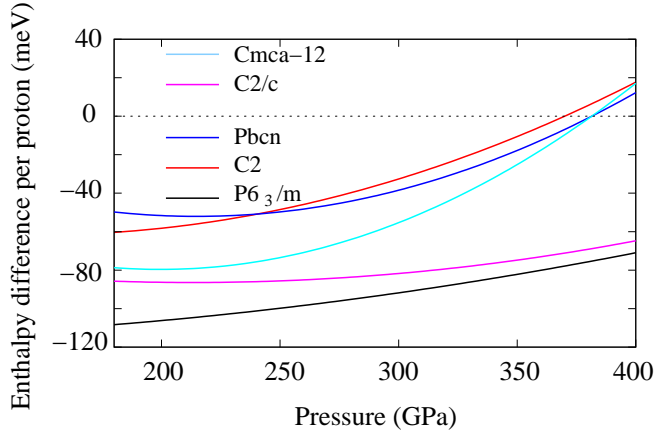}
\caption{
\label{enthalpy-pbe0}
(Colors online) The enthalpy per proton as a function of pressure with respect to the metallic Cmca structure at the hybrid DFT level of theory.} 
\end{figure}
Indeed, looking at Fig.~\ref{enthalpy-pbe0} a completely altered phase diagram can be observed that energetically favors all insulating phases over the metallic Cmca structure. The reduction in relative enthalpy of the hexagonal P6$_{3}$/m phase, which is not competitive at all at the semilocal DFT level, is particularly dramatic and eventually renders the structure stable over the whole pressure range of Phase III. Our calculations suggest that this can attributed to the lower pressure estimate at the hybrid DFT level. In any event, contrary to earlier theoretical predictions \cite{2000Natur.403..632J}, the metallic Cmca structure can be definitely ruled out at this point. Even though, it had been recently recognized that Cmca may only become stable at very high compression $\sim$400 GPa \cite{PhysRevLett.84.6070, 2007NatPh...3..473P}, we find that irrespective of zero-point energy (ZPE), which will be discussed later,  it is never stable. Another qualitative difference to previous calculations at the semilocal DFT level is, that relative to the Cmca-12 structure, the C2/c phase is throughout more stable over the whole pressure range. Even more, the latter is even further stabilized with pressure and leaves the C2/c structure as the only other potential candidate for Phase III. Although the much increased stability of the C2/c and of the even more favorable P6$_{3}$/m structure may at first sight come as a surprise, both are consequences of the aforementioned energy lowering of insulating phases within hybrid DFT.
\begin{figure}
\centering
\includegraphics[width=1.0\columnwidth]{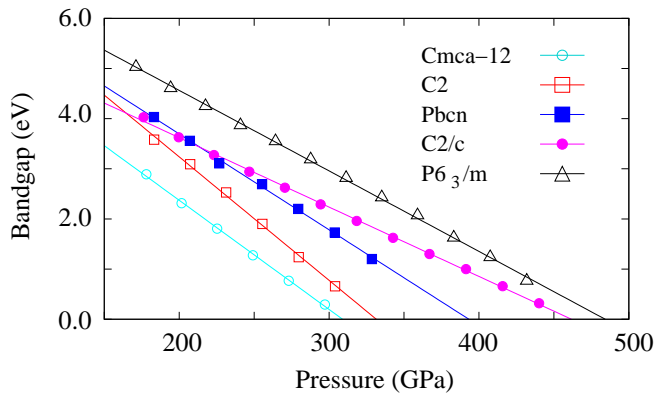}
\caption{
\label{PBE0-gap}
(Colors online) The single-particle bandgap for all considered crystal structures as functions of pressure at the hybrid DFT level.}
\end{figure}
Furthermore, the impact of hybrid DFT on the bandgap can be seen in Fig.~\ref{PBE0-gap}. It is apparent that in comparison to the semilocal DFT gap of Fig.~\ref{PBE-gap}, the bandgaps for all considered phases and thus the corresponding metallization pressures are substantially increased. In conjunction with Fig.~\ref{enthalpy-pbe0}, we find that the most stable structure P6$_{3}$/m exhibits at the same time the largest bandgap, which is at variance to the semilocal DFT results and eventually entails an even higher metallization pressure. In general, except for the Cmca-12 structure, which up to $\sim$375 GPa is more stable than expected, the aforementioned rule of Kaxiras et al. is very well maintained at the hybrid DFT level.

In particular, we report that at the hybrid DFT level the paired but insulating state in the form of the P6$_{3}$/m phase is insistent and stable up to $\sim$500 GPa, with no transformation into a metallic pairing structure expected. In addition, metallization within P6$_{3}$/m via bandgap closure will only occur for pressures higher than 484 GPa. That is to say that the combination of HFX properties to widen the bandgap, while stabilizing insulating phases, which from the outset themselves exhibit wider gaps, is responsible for the tremendous increase of metallization pressure with respect to semilocal DFT. Obviously, the quantitative increase of the metal-insulator transition pressure from $\sim$290 GPa to 484 GPa when going from semilocal to hybrid DFT is a consequence of the qualitative change in the mechanism of metallization. While in the former case metallization happens at $\sim$290 GPa by bandgap closure of the most favorable C2/c structure, where it concurrently transforms to the metallic Cmca phase, now P6$_{3}$/m is stable up to the highest pressures and metallizes at 484 GPa. Preliminary calculations using more accurate many-body perturbation theory, which will be reported in detail elsewhere, confirm the relative ordering of the bandgap calculations as shown in Fig.~\ref{PBE0-gap} for all considered structures. Quantitatively, for the small bandgap structures such as Cmca-12 the metallization pressures insignificantly increases, while for the C2/c and P6$_{3}$/m phases it marginally decreases. Anyway, in all cases the deviation in the transition pressure is smaller than 60 GPa and we eventually predict that the most relevant P6$_{3}$/m structure metallizes due to bandgap closure at 446 GPa. 
Since this is in excellent agreement with the experimental observation that solid hydrogen turns opaque at 320 GPa, from which a metallization pressure of 450 GPa can be deduced \cite{2002Natur.416..613L}, our calculations resolve one of the primarily inconsistencies between theory \cite{2000Natur.403..632J, PhysRevLett.84.6070, 2007NatPh...3..473P} and experiment \cite{PhysRevLett.76.1663, PhysRevLett.76.1667}. 

The pressure upon which solid molecular hydrogen dissociates has been recently theoretically determined to be 490 GPa \cite{PhysRevLett.106.165302}, which immediately suggest that Phase III may never become metallic through bandgap closure, but rather directly transforms into the metallic atomic phase. As a consequence, we do not expect any electron-phonon driven superconductivity in solid molecular hydrogen, which is in contrast to previous previous theoretical predictions \cite{1989Natur.340..369B, PhysRevLett.78.118}. Even though the transition pressure has only been computed at the semilocal DFT level, as the coexistence of the Cmca and the I4$_{1}$/amd phase, it can anyhow be considered as rather reliable, since both structures are metallic so that no appreciable corrections due to HFX are expected. Nevertheless, the fact that we predict P6$_{3}$/m to be substantially more stable than the Cmca phase, would involve a slightly higher dissociation pressure. On the other hand, we also expect the inclusion of ZPE not only to increase the transition pressure by $\sim$10\%, but alike to somewhat increase the bandgap because of level-repulsion effects, which increases the metallization pressure as well \cite{PhysRevLett.84.6070}. Including all these effects, we expect the metallization to increase above 490 GPa and Phase III to be insulating until it dissociates into metallic atomic hydrogen. 

Speaking about ZPE, the remaining factor is its influence on the relative stability of different structures. However, when comparing Fig.~\ref{enthalpy-pbe} with Fig.~\ref{enthalpy-pbe0} the stabilization of HFX is such that just the earlier predicted C2/c phase \cite{2007NatPh...3..473P} appears to be the only other competitive structure. Although in absolute value the ZPE is rather large, contrary to solid atomic hydrogen, there is a strong tendency in the molecular case that the ZPE of different structures cancel each other out, which is even more pronounced with increasing pressure \cite{PhysRevLett.74.1601}. 
In any case, the fact that the ZPE tends to favor symmetric structures \cite{PhysRevB.36.2092, PhysRevLett.38.415} further strengthens our prediction in favor of the P6$_{3}$/m phase as the most likely candidate for Phase III. 

All of this can be understood in terms of the concept of spontaneous polarization \cite{1997Natur.388..652E}. The fact that upon enthalpy minimization the hydrogen molecules are allowed to slightly move away from their ideal lattice sites has several important implications. First of all it stabilizes the P6$_{3}$/m phase, where the centers of the hydrogen molecules sit on a distorted hcp lattice. Moreover, it also entails a substantial asymmetric electronic charge distribution, which causes the system to spontaneously polarize. This electronic symmetry-breaking in the proton pairs not only accounts for the established bandgap widening, thereby yielding the much increased metallization pressure, but also explains the existence of IR active vibron modes \cite{1997Natur.388..652E}. In fact, it had been recently shown that P6$_{3}$/m obeys one intense IR active vibron mode \cite{2007NatPh...3..473P}, which is consistent with experimental IR measurements of Phase III \cite{1994Natur.369..384H}. Altogether, we conclude that P6$_{3}$/m is a very likely candidate for the elusive Phase III of solid hydrogen. 

Having said that, in spite the broken symmetry phase is believed to exhibit a distorted hcp lattice because of the continuous shift of the Raman and IR frequencies next to the Phase II-III transition, the suggested P6$_{3}$/m structure is inconsistent with Phase II that has three active vibron modes with an altogether low IR activity and therefore less symmetry \cite{2007NatPh...3..473P, PhysRevLett.78.2783}. Anyhow, not only the level of theory may be inadequate, but also the enthalpy differences so small that entropic effects are no longer negligible and a finite temperature treatment essential \cite{PhysRevLett.98.066401}. 

We conclude by noting that the predominance of the proposed P6$_{3}$/m phase delimits the pressure range of existence of the conjectured zero-temperature quantum liquid phase of metallic hydrogen \cite{2004Natur.431..669B}.

\begin{acknowledgments}
The authors would like to thank the Graduate School of Excellence MAINZ and the IDEE project of the Carl Zeiss Foundation for financial support, as well as Hans Behringer for critically reading the manuscript.
\end{acknowledgments}

\bibliography{tdk} 

\end{document}